\documentclass[reprint,amsmath,amssymb,aps,prd]{revtex4-2}
\usepackage{bm}
\usepackage{dcolumn}
\usepackage{lineno}
\usepackage{graphicx}
\usepackage{subfigure}
\usepackage{epstopdf}
\usepackage{float}
\usepackage{threeparttable}
\usepackage{overpic}
\usepackage{subfigure}
\usepackage{float}
\usepackage{ulem}
\usepackage{cmap} 
\usepackage{mathtext}
\usepackage{mathrsfs}
\usepackage{multirow}
\usepackage{xcolor}
\definecolor{aa}{RGB}{0,0,139}
\usepackage[bookmarksnumbered=true,colorlinks,urlcolor=aa,linkcolor=blue,anchorcolor=aa,citecolor=aa]{hyperref}

\newcommand{\chicJ}{\chi_{cJ}}

\newcommand{\bfg}{\begin{figure}}
\newcommand{\efg}{\end{figure}}
\newcommand{\bitm}{\begin{itemize}}
\newcommand{\eitm}{\end{itemize}}
\newcommand{\bnum}{\begin{enumerate}}
\newcommand{\enum}{\end{enumerate}}
\newcommand{\btbl}{\begin{table}}
\newcommand{\etbl}{\end{table}}
\newcommand{\btbu}{\begin{tabular}}
\newcommand{\etbu}{\end{tabular}}
\newcommand{\bcl}{\begin{center}}
\newcommand{\ecl}{\end{center}}

\newcommand{\beq}{\begin{equation}}
\newcommand{\eeq}{\end{equation}}
\newcommand{\beqr}{\begin{eqnarray}}
\newcommand{\eeqr}{\end{eqnarray}}

\definecolor{boslv}{rgb}{0.0, 0.65, 0.58}
\definecolor{Munsell}{HTML}{00A877}

\newcommand{\Br}{\mathcal{B}}

\begin{document}
\title{\boldmath Observation of
  $\chi_{c0}\to\Sigma^{+}\bar{\Sigma}^{-}\eta$ and evidence for
  $\chi_{c1,2}\to\Sigma^{+}\bar{\Sigma}^{-}\eta$}

\author{
\begin{small}
\begin{center}
M.~Ablikim$^{1}$, M.~N.~Achasov$^{4,c}$, P.~Adlarson$^{75}$, O.~Afedulidis$^{3}$, X.~C.~Ai$^{80}$, R.~Aliberti$^{35}$, A.~Amoroso$^{74A,74C}$, Q.~An$^{71,58,a}$, Y.~Bai$^{57}$, O.~Bakina$^{36}$
% ...
\\
\vspace{0.2cm}
(BESIII Collaboration)\\
\vspace{0.2cm} {\it
$^{1}$ Institute of High Energy Physics, Beijing 100049, People's Republic of China\\
% ...
$^{3}$ Bochum  Ruhr-University, D-44780 Bochum, Germany\\
$^{4}$ Budker Institute of Nuclear Physics SB RAS (BINP), Novosibirsk 630090, Russia\\
% ...
$^{35}$ Johannes Gutenberg University of Mainz, Johann-Joachim-Becher-Weg 45, D-55099 Mainz, Germany\\
$^{36}$ Joint Institute for Nuclear Research, 141980 Dubna, Moscow region, Russia\\
% ...
$^{57}$ Southeast University, Nanjing 211100, People's Republic of China\\
$^{58}$ State Key Laboratory of Particle Detection and Electronics, Beijing 100049, Hefei 230026, People's Republic of China\\
% ...
$^{71}$ University of Science and Technology of China, Hefei 230026, People's Republic of China\\
% ...
$^{74}$ University of Turin and INFN, (A)University of Turin, I-10125, Turin, Italy; (B)University of Eastern Piedmont, I-15121, Alessandria, Italy; (C)INFN, I-10125, Turin, Italy\\
$^{75}$ Uppsala University, Box 516, SE-75120 Uppsala, Sweden\\
% ...
$^{80}$ Zhengzhou University, Zhengzhou 450001, People's Republic of China\\
\vspace{0.2cm}
$^{a}$ Deceased\\
% ...
$^{c}$ Also at the Novosibirsk State University, Novosibirsk, 630090, Russia\\
% ...
}
\end{center}
\vspace{0.4cm}
\end{small}
}

\date{\today}

%

\begin{abstract}

Using $(27.12\pm 0.14)\times10^{8}$ $\psi(3686)$ events collected with
the BESIII detector, the decay 
$\chi_{c0}\to\Sigma^{+}\bar{\Sigma}^{-}\eta$ is observed for the first
time with a statistical significance of $7.0\sigma$, and evidence for
$\chi_{c1}\to\Sigma^{+}\bar{\Sigma}^{-}\eta$ and
$\chi_{c2}\to\Sigma^{+}\bar{\Sigma}^{-}\eta$ is found with
statistical significances of $4.3\sigma$ and $4.6\sigma$,
respectively.
The branching fractions are determined to be
$\Br(\chi_{c0}\to\Sigma^{+}\bar{\Sigma}^{-}\eta)=({1.26 \pm 0.20 \pm
  0.13}) \times 10^{-4},
~\Br(\chi_{c1}\to\Sigma^{+}\bar{\Sigma}^{-}\eta)=({5.10 \pm 1.21 \pm
  0.67}) \times 10^{-5}$, and
$\Br(\chi_{c2}\to\Sigma^{+}\bar{\Sigma}^{-}\eta)=({5.46 \pm 1.18 \pm
  0.50}) \times 10^{-5}$, where the first uncertainties are
statistical, and the second ones are systematic.

\end{abstract}

\maketitle

\section{Introduction}\label{sec:introduction}

In the investigations of $e^+e^-\to\Lambda\bar{\Lambda}\eta(\phi)$ and
$J/\psi(\psi(3686))\to\gamma p \bar{p}$, unexpected enhancements have
been detected near the mass thresholds of $\Lambda\bar{\Lambda}$ and
$p\bar{p}$ pairs~\cite{xuwei, enhance_LLP, yanpingH}. Several
theoretical models have been suggested to explain these enhancements,
including the one-boson-exchange potential model, $^{3}P_{0}$ meson
decay model, quark potential model and quark-pair creation
model~\cite{enhancement1, enhancement2}.  Due to a larger phase space
available for $\chi_{cJ}$ decays compared to $J/\psi$ decays, a
greater range of possible final states may be explored in $\chi_{cJ}$
decays for baryon--anti-baryon pair mass threshold enhancements.

The exploration of charmonium decays is crucial to improve our
understanding of Quantum Chromodynamics
(QCD)~\cite{Brambilla:2010cs}. So far, only a few investigations have
been conducted on the decays $\chi_{cJ}\to B\bar BM$ (where $B$
represents a baryon and $M$ denotes a meson), such as
$\chi_{cJ}\to\Lambda\bar{\Lambda}\eta$~\cite{zyj}. Therefore, further
studies are highly desirable to explore the properties of
$\chi_{cJ}$ particles.

In this paper, we report the first observation of
$\chi_{c0}\to\Sigma^{+}\bar{\Sigma}^{-}\eta$, evidence for
$\chi_{c1,2}\to\Sigma^{+}\bar{\Sigma}^{-}\eta$, and searches for
enhancements near the $\Sigma^{+}\bar{\Sigma}^{-}$ mass threshold and possible
excited states of $\Sigma^{+}$, where the $\chi_{c0,1,2}$ are produced
in $\psi(3683)$ radiative decays.  The analysis uses $(27.12\pm 0.14)\times
10^{8}$ $\psi(3686)$ events~\cite{Ablikim:2017wyh} collected with the
BESIII detector in 2009, 2012 and 2021.

\section{BESIII Detector and Monte Carlo Simulation}\label{sec:simu}

The BESIII detector~\cite{Ablikim:2009aa} records symmetric $e^+e^-$
collisions provided by the BEPCII storage
ring~\cite{Yu:IPAC2016-TUYA01} in the center-of-mass energy range from
2.0 to 4.95~GeV, with a peak luminosity of $1 \times
10^{33}\;\text{cm}^{-2}\text{s}^{-1}$ achieved at $\sqrt{s} =
3.77\;\text{GeV}$. BESIII has collected large data samples in this
energy region~\cite{Ablikim:2019hff, EcmsMea, EventFilter}. The
cylindrical core of the BESIII detector covers 93\% of the full solid
angle and consists of a helium-based multilayer drift chamber~(MDC), a
plastic scintillator time-of-flight system~(TOF), and a CsI(Tl)
electromagnetic calorimeter~(EMC), which are all enclosed in a
superconducting solenoidal magnet providing a 1.0~T magnetic
field. The solenoid is supported by an octagonal flux-return yoke with
resistive plate counter muon identification modules interleaved with
steel. The charged-particle momentum resolution at $1~{\rm GeV}/c$ is
$0.5\%$, and the specific energy loss (${\rm d}E/{\rm d}x$) resolution is $6\%$ for electrons
from Bhabha scattering. The EMC measures photon energies with a
resolution of $2.5\%$ ($5\%$) at $1$~GeV in the barrel (end cap)
region. The time resolution in the TOF barrel region is 68~ps, while
that in the end cap region was 110~ps. The end cap TOF system was
upgraded in 2015 using multigap resistive plate chamber technology,
providing a time resolution of 60~ps, which benefits 86\% of the data
used in this analysis~\cite{etof}.

Monte Carlo (MC) simulated data samples produced with a
{\sc geant4}-based~\cite{geant4} software package, which includes the geometric description of the BESIII detector and the detector response, are used to determine detection efficiencies and to estimate backgrounds. The simulation models the beam energy spread and initial state radiation (ISR) in the $e^+e^-$ annihilations with the generator {\sc kkmc}~\cite{ref:kkmc}. The inclusive MC
    sample includes the production of the $\psi(3686)$ resonance, the ISR production of the $J/\psi$, and the continuum processes incorporated in {\sc kkmc}, with approximately $~2.7$ billion events. All particle decays are modeled with
    {\sc evtgen}~\cite{ref:evtgen} using branching fractions either
    taken from the Particle Data Group (PDG)~\cite{pdg}, when
    available, or otherwise estimated with {\sc
      lundcharm}~\cite{ref:lundcharm}. In this analysis, in order to
    take possible intermediate structures into consideration, we use
    the BODY3~\cite{BODY3} model to generate signal MC events~(3 million events for each channel), by
    reweighting the phase space (PHSP) MC Dalitz distribution to match
    the background-subtracted data.

\section{Event Selection}\label{sec:selection}

The $\Sigma^{+}(\bar{\Sigma}^{-})$ candidate is reconstructed via
$\Sigma^{+}(\bar{\Sigma}^{-}) \to p \pi^{0}(\bar{p}\pi^{0})$, and the
$\eta$ candidate is reconstructed via $\eta\to \gamma\gamma$.

Photon candidates are identified using isolated showers in the EMC.  The
deposited energy of each shower must be more than 25~MeV in the barrel
region ($|\cos \theta|< 0.80$), where $\theta$ is the polar angle
with respect to the $z$ axis, which is the symmetry axis of
the MDC, and more than 50~MeV in the end cap region ($0.86 <|\cos
\theta|< 0.92$). To exclude showers that originate from charged
tracks, the angle subtended by the EMC shower and the position of the
closest charged track at the EMC must be greater than 10 degrees as
measured from the interaction point (IP). To suppress electronic noise
and showers unrelated to the event, the difference between the EMC
shower time and the event start time is required to be within [0,
  700]\,ns. The number of photon candidates is required to be at least
seven.

Candidate events must contain at least one positively charged track
and one negatively charged track. The polar angle of each
track measured in the MDC is required to satisfy
${\left|\cos\theta\right|} < 0.93$. The d$E/$d$x$ information in the
MDC is combined with the time of flight from the TOF detector to
identify the type of particle (PID). For this purpose, confidence
levels for pion, proton and kaon hypotheses are calculated, and tracks
are assigned to the hypothesis with the highest confidence level. The
$\Sigma^{+}(\bar{\Sigma}^{-})$ candidate is reconstructed by combining
pairs of $p(\bar{p})$ and $\pi^{0}$ candidates. Because of the
relatively large decay length of the $\Sigma^{+}(\bar{\Sigma}^{-})$
particle, the $p(\bar{p})$ candidate is required to have a distance
of closest approach to the IP less than 15 cm along the $z$-axis, and
less than 2 cm in the transverse plane. 

Next we select $\pi^0$ candidates from $\gamma\gamma$ pairs with
invariant masses within [0.08, 0.20]~GeV$/c^2$.  A $\pi^0$ candidate
must satisfy a one constraint (1C) kinematic fit with the
$\gamma\gamma$ mass constrained to the $\pi^0$ mass. At least two
$\pi^0$ candidates are required.  Of the remaining photons, the one
resulting in the largest value of $\Delta M$ as defined in Eq.~\ref{eq1} is assigned as
the radiative photon, and the others are assigned as possibly decaying
from the $\eta$.

In order to suppress background and improve the resolution,
six-constraint (6C) kinematic fits (with constraints on total four
momentum of the final state particles and the masses of the two
$\pi^0$ mass) are applied to all the potential final state
combinations, and the combination with the smallest $\chi_{6\rm
  C}^{2}$ is kept.  Further, to match $p$($\bar{p}$) with its
$\pi^0$ to identify $\Sigma^{+}(\bar{\Sigma}^{-})$,
only the combination with the smallest
value of
\begin{eqnarray}
  \Delta M &=& \left((M_{p\pi^0} - m_{\Sigma^+})^2
               + (M_{\bar p \pi^0} - m_{\bar\Sigma ^-})^2
               \right.
               \nonumber \\
           && + \left. (M_{\gamma\gamma}-m_{\eta})^2 \right)^{\frac{1}{2}}
              \label{eq1}
\end{eqnarray}
is retained, where $M_{p\pi^0}(M_{\bar p \pi^0})$ is the invariant
mass of the $p\pi^0(\bar p \pi^0)$ system, $M_{\gamma\gamma}$ is the
invariant mass of the $\gamma\gamma$ system, $M_{\Sigma^{+}}(M_{\bar
  \Sigma^-})$ is the mass of $\Sigma^{+}(\bar\Sigma^-)$~\cite{pdg},
and $M_{\eta}$ is the mass of $\eta$~\cite{pdg}.  A figure of merit
(FOM) optimization is performed for the $\chi_{6\rm C}^{2}$
requirement, based on maximizing the value of $\frac{S}{\sqrt{S+B}}$
(where $S$
is the signal yield from the signal MC sample and $B$ is the
background yield from the inclusive MC sample normalized to the
integrated luminosity of data). The optimized selection criterion is
$\chi_{6\rm C}^{2}<45$.

Various background channels are suppressed by
requiring $\chi_{\rm signal}^2 < \chi_{\rm bkg}^2$, where $\chi_{\rm
  signal}^2$ is the 4C $\chi^{2}$ under the hypothesis of
$\chi_{cJ}\to\Sigma^{+}\bar{\Sigma}^{-}\eta$, while $\chi_{\rm
  bkg}^{2}$ is the 4C $\chi^{2}$ of background
channels by adding or subtracting one photon, satisfying the good
photon requirements.

In the following, particle momenta updated by the 6C fit are used.  To reject
$\chi_{cJ}\to\gamma J/\psi(\to\Sigma^{+}\bar{\Sigma}^{-}\gamma)$, we
require $|M(\Sigma^{+}\bar{\Sigma}^{-}\gamma_{E})-3.091~{\rm
  GeV}/c^{2}| > 0.056~$GeV$/c^{2}$, where $\gamma_{E}$ is the photon
from the $\psi(3686)$. To reject
$\chi_{cJ}\to\Sigma^{+}\bar{\Sigma}^{-}\pi^{0}$, we require
$|M(\gamma_{E}\gamma_{1})-0.132~{\rm GeV}/c^{2}|> 0.018~$GeV$/c^{2}$
and $|M(\gamma_{E}\gamma_{2})-0.137~{\rm GeV}/c^{2}|>0.044~{\rm
  GeV}/c^{2}$, where $\gamma_{1}$ and $\gamma_{2}$ are the photons
from the $\eta$. The mass windows for these vetoes are centered around
the fitted value from data, which slightly differ from the PDG
values. The widths of the mass windows are also obtained from the fit
to data and correspond to the $3~\sigma$ region around the fitted central value. Other potential backgrounds, including $\chi_{cJ}\to
\eta \Delta^{+}\bar{\Delta}^{-}, \chi_{cJ}\to \eta^{\prime}p\bar{p}$
and $\chi_{cJ}\to \pi^{0}\Sigma^{+} \bar{\Sigma}^{-}$, are
investigated by analyzing the $\psi(3686)$ inclusive MC sample with
TopoAna~\cite{TopoAna}. After the above requirements, only a small
fraction (less than 1\% of the signal yield) of background events
survive the selection criteria and can be safely ignored.  We impose
the same event selection criteria on the continuum data taken at
$\sqrt{s}= 3.650$~GeV, and only one event survives. After scaling to
the $\psi(3686)$ data sample using $f_{\rm{scale}} =
{{\mathcal{L}_{\psi(3686)}}\over{\mathcal{L}_{3.65}}
}\times{{3.65^2}\over{3.686^2}}$, where the $\mathcal{L}_{\psi(3686)}$
and $\mathcal{L}_{3.65}$ are the luminosities of the $\psi(3686)$ data
and continuum data samples, respectively, the
continuum contribution is also found to be negligible.

\begin{figure*}[htbp]
\begin{center}
\begin{minipage}[t]{0.445\linewidth}
\includegraphics[width=1.0\textwidth]{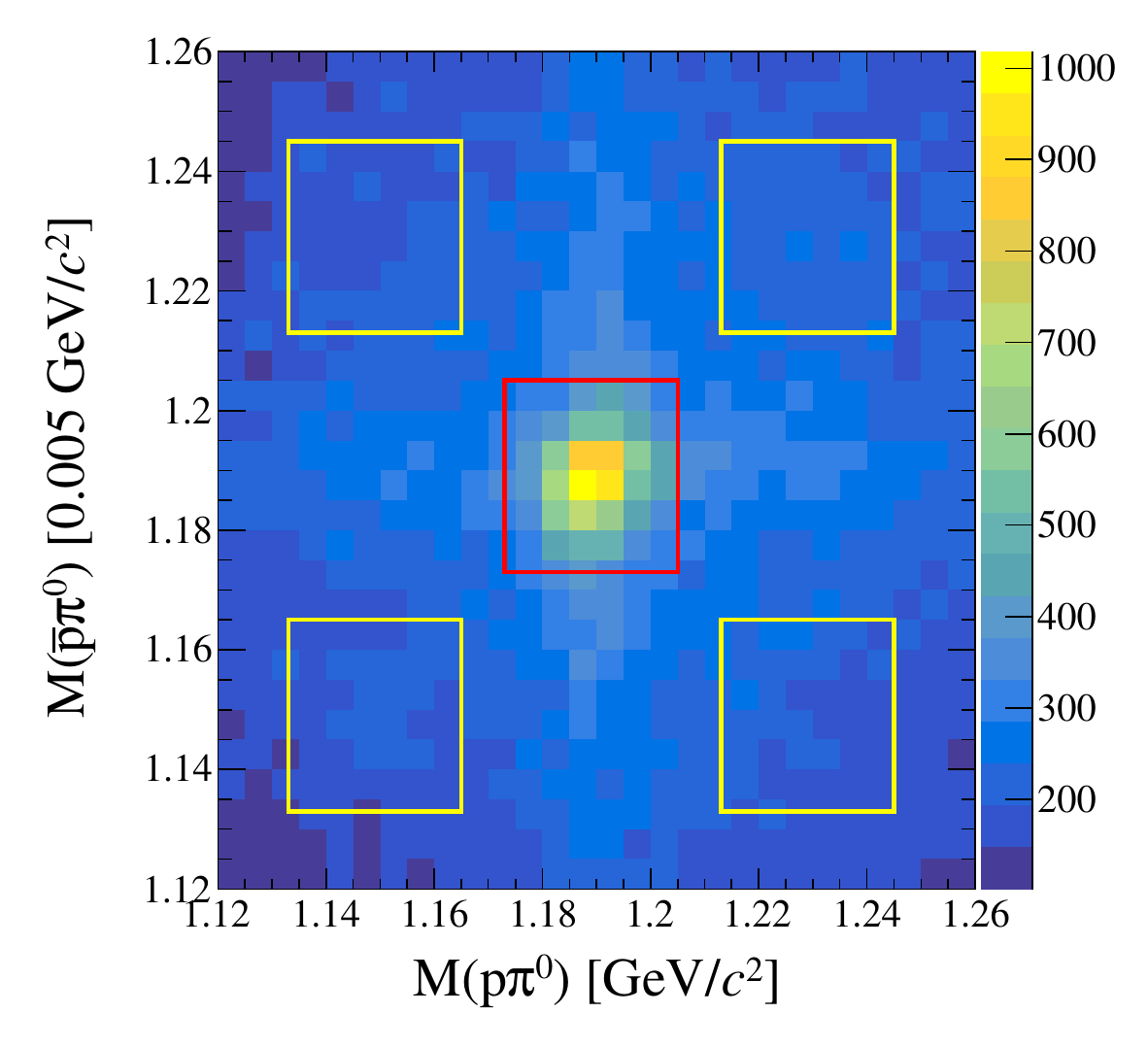}
\end{minipage}%
\begin{minipage}[t]{0.55\linewidth}
\includegraphics[width=1.0\textwidth]{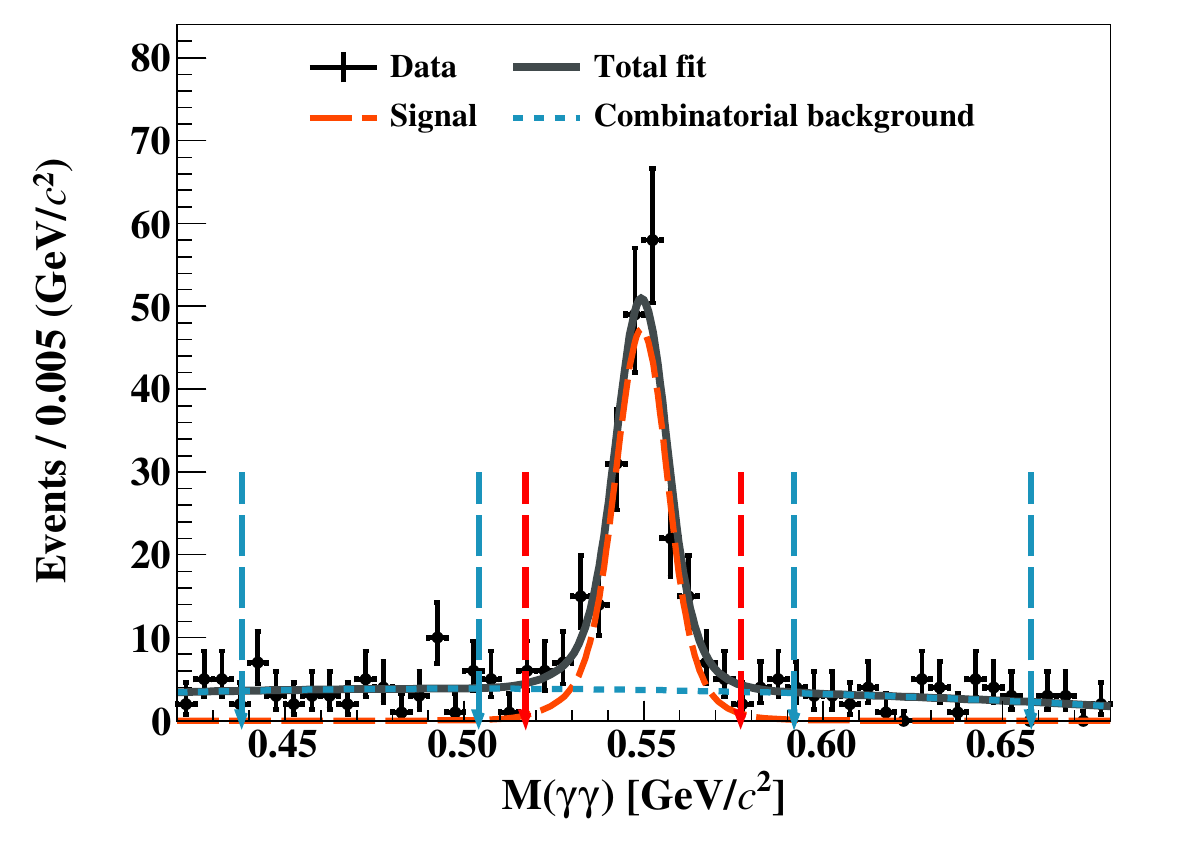}
\end{minipage}%

\caption{Distributions of (left) $M(\bar p \pi^0)$ versus $M(p
  \pi^0)$ and (right) $M(\gamma\gamma)$ of the accepted candidates. In
  the left figure, the red box represents the $\Sigma^{+}\bar
  \Sigma^{-}$ signal region, and the yellow boxes are the $\Sigma^{+}\bar
  \Sigma^{-}$ sideband region. In the right figure, the red dashed
  line represents the fitted $\eta$ signal, and the blue dashed line
  is the combinatorial background. The grey line is the total
  fit. The red arrows denote the $\eta$ signal region, while the blue
  arrows denote the $\eta$ sideband region.}

\label{fig:ConSam_pppipi_scatter_2Lambda}
\end{center}
\end{figure*}

The distributions of $M(\bar p \pi^0)$ versus $M(p \pi^0)$ and
$M(\gamma\gamma)$ with all the selection criteria are shown in
Fig.~\ref{fig:ConSam_pppipi_scatter_2Lambda}.  The $\Sigma^{+}/\bar
\Sigma^{-}$ signal mass window of $M{(p\pi)}$ is chosen as $[1.174,
  1.204]$ GeV/$c^{2}$, and the one dimensional (1D) sideband regions
are $[1.139, 1.169]$ GeV/$c^{2}$ and $[1.209, 1.239]$~GeV/$c^{2}$. The
four squares with equal areas around the signal region are taken as
the two dimensional (2D) $\Sigma^{+}\bar \Sigma^{-}$sideband
regions. The $\eta$ signal mass window is taken as $M(\gamma\gamma)$
$\in$ $[0.517, 0.577]~\rm GeV/c^{2}$, and the $\eta$ sideband regions
are $[0.448, 0.508]$ GeV/$c^{2}$ and $[0.588, 0.648]$~GeV/$c^{2}$.

\section{Signal yield determination}\label{Sec:sig}

\begin{figure*}[htbp]
  \begin{center}
  \begin{minipage}[t]{1.0 \linewidth}
  \includegraphics[width=1\textwidth]{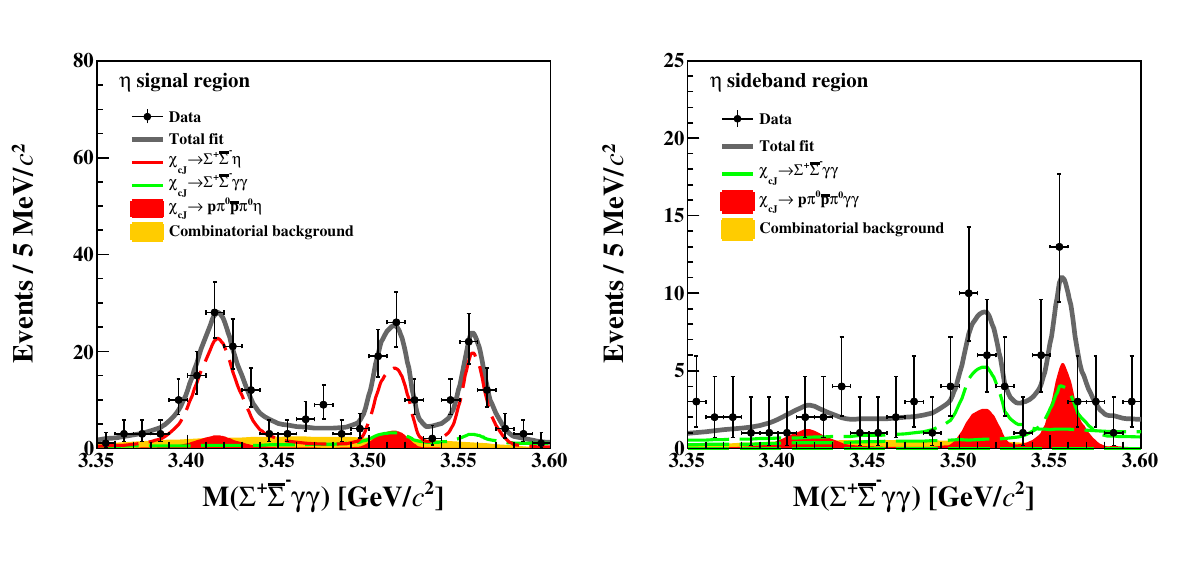}
  \end{minipage}
\caption{ Simultaneous fit to the
  $M(\Sigma^+\bar{\Sigma}^-\gamma\gamma)$ distributions in the $\eta$
  signal (left) and sideband (right) regions. In the left figure, the
  red dashed line is the signal, the red histogram is the fixed
  contribution from the non-$\Sigma^+\bar{\Sigma}^-$ sideband region
  of data in the $\eta$ signal region, and the green dashed line is
  the background contribution constrained by the fit to the $\eta$
  sideband. Additionally, the orange histogram is the combinatorial
  background. The grey line is the total fit. In the right figure, the
  green line is from the simulated $\chi_{cJ} \to
  \Sigma^+\bar{\Sigma}^-\gamma\gamma$ shape, and the red histogram is
  the fixed contribution from the non-$\Sigma^+\bar{\Sigma}^-$
  background estimated by the $\Sigma^+\bar{\Sigma}^-$ sideband region
  of data.  }
\label{fit_result}
\end{center}
\end{figure*}

The signal yields of $\chi_{cJ}$ decays are determined by performing a
simultaneous fit to the $M(\Sigma^+\bar{\Sigma}^-\gamma\gamma)$
distributions for events in both the $\eta$ signal and sideband
regions, as shown in Fig.~\ref{fit_result}.

For the fit to the events in the $\eta$ signal region, shown in
Fig.~\ref{fit_result} (left), the probability density functions of the
$\chi_{cJ}$ signals are modeled by individual simulated MC shapes
convolved with a Gaussian resolution function that accounts for the
resolution difference between data and MC simulation. The
combinatorial background is described by a second order Chebyshev
polynomial function. The non-$\eta$ background is constrained by the
simultaneous fit to the events in the $\eta$ sideband region. The
non-$\Sigma^{+}\bar{\Sigma}^{-}$ background is fixed to the number
obtained from the non-$\Sigma^{+}\bar{\Sigma}^{-}$ background
estimated by the $\Sigma^{+}\bar{\Sigma}^{-}$ sideband region of data
as shown in Fig.~\ref{fig:ConSam_pppipi_scatter_2Lambda} (right) with
a scale factor of 0.25, since the $\Sigma^{+}\bar{\Sigma}^{-}$
sideband region is four times larger than the
$\Sigma^{+}\bar{\Sigma}^{-}$ signal region.

For the fit to the $\eta$ sideband region, the same probability
density functions are used for the $\chi_{cJ}$ signal shapes.  The
combinatorial background is described by a second order Chebyshev
polynomial function.  The non-$\Sigma^{+}\bar{\Sigma}^{-}$ background
is again constrained by the number of events in the
$\Sigma^{+}\bar{\Sigma}^{-}$ sideband region with a scale factor of
0.25. To determine the scale factor between the $\eta$ signal and
sideband regions, $f_{\eta}$, a fit is performed on the
$M(\gamma\gamma)$ spectrum, as shown in
Fig.~\ref{fig:ConSam_pppipi_scatter_2Lambda} (right), in which the
simulated signal MC shape convolved with a Gaussian function is used
to model the $\eta$ signal and a polynomial function is used to
describe the combinatorial background. The scale factor $f_{\eta}$ is
determined to be 0.547.

For the above two fits, the number of non-$\Sigma^+\bar{\Sigma}^-$
background events in the $\Sigma^+\bar{\Sigma}^-$ sideband region
(denoted as the four yellow boxes in
Fig. \ref{fig:ConSam_pppipi_scatter_2Lambda}(left)) are $11.4\%$ in
the $\eta$ signal region and $37.9\%$ in the $\eta$ sideband region.
The signal yields for $\chi_{c0,1,2}\to \Sigma^+\bar{\Sigma}^-\eta$
are $74\pm 12, 36\pm 8$ and $35\pm 8$, with statistical significances
of $7.0\sigma, 4.3\sigma$, and $4.6\sigma$, respectively. The
statistical significance is determined by examining the difference in
log-likelihood as each signal is individually excluded in the fit,
taking the changes in the degrees of freedom into account.

The branching fractions of $\chi_{cJ}\to \Sigma^+\bar{\Sigma}^-\eta$
are calculated by

\begin{equation}
\Br({\chi _{cJ}} \to \Sigma^+\bar{\Sigma}^-\eta ) = {\textstyle{{N_{\rm fit} \over {N_{\psi(3686)} \cdot \Br\left( {\psi(3686) \to \gamma {\chi _{cJ}}} \right)\cdot {\prod _i}  {\Br_i} \cdot \epsilon }}}},
\end{equation}
where $N_{\rm fit}$ is the fitted signal yield of $\chicJ$,
${N_{\psi(3686)}}$ is the number of $\psi(3686)$ events,
${\mathcal B}(\psi(3686)\to\gamma\chi_{cJ})$ is the branching
fractions of $\psi(3696)\to\gamma\chi_{cJ}$, ${\prod _i} {\Br_i}$ is
the product of branching fractions of the intermediate decays,
including $\Br(\Sigma^+ \rightarrow p \pi^{0})=(51.57\pm0.30)\%$,
$\Br(\bar{\Sigma}^- \rightarrow \bar{p} \pi^{0})=(51.57\pm0.30)\%$,
$\Br(\pi^{0} \rightarrow \gamma \gamma)=(98.823 \pm 0.034)\%$ and
$\Br(\eta \rightarrow \gamma\gamma)=(39.36\pm0.18)\%$, which are taken
from the PDG~\cite{pdg}, and $\epsilon$ is the detection
efficiency. The results of branching fractions are listed in
Table~\ref{list:Br}.

\begin {table}[htbp]
\begin{center}
\fontsize{8}{10}\selectfont

\renewcommand\arraystretch{1.2}

{\caption {Fitted signal yield~($N_{\rm fit}$), detection
    efficiency ($\epsilon$), statistical significance, and
    branching fraction ($\mathcal B$). The first and second
    uncertainties are statistical and systematic, respectively.  }
\label{list:Br}}
\begin{tabular}{c|c|c|c|c}
 \hline \hline
 Decay   & $N{_{\rm fit}}$  & Significance &  $\epsilon~(\%)$ & $\mathcal B~(10^{-5})$ \\   \hline
  
  $\chi_{c 0} \rightarrow \Sigma^+\bar{\Sigma}^-\eta$ & $74\pm 12$ & $7.0\sigma$ & 2.18  & $12.6\pm2.0\pm1.3$ \\
   
 $\chi_{c 1} \rightarrow \Sigma^+\bar{\Sigma}^-\eta$  & $36\pm 8$  & $4.3\sigma$ & 2.61  &  $5.10 \pm 1.21 \pm 0.67$ \\

 $\chi_{c 2} \rightarrow \Sigma^+\bar{\Sigma}^-\eta$  & $35\pm 8$  & $4.6\sigma$ & 2.46  &   $5.46 \pm 1.18 \pm 0.50$ \\  \hline\hline

\end{tabular}
\end{center}
\end{table}

\begin{figure*}[htbp]
\begin{center}
\begin{minipage}[t]{0.49\linewidth}
\includegraphics[width=1\textwidth]{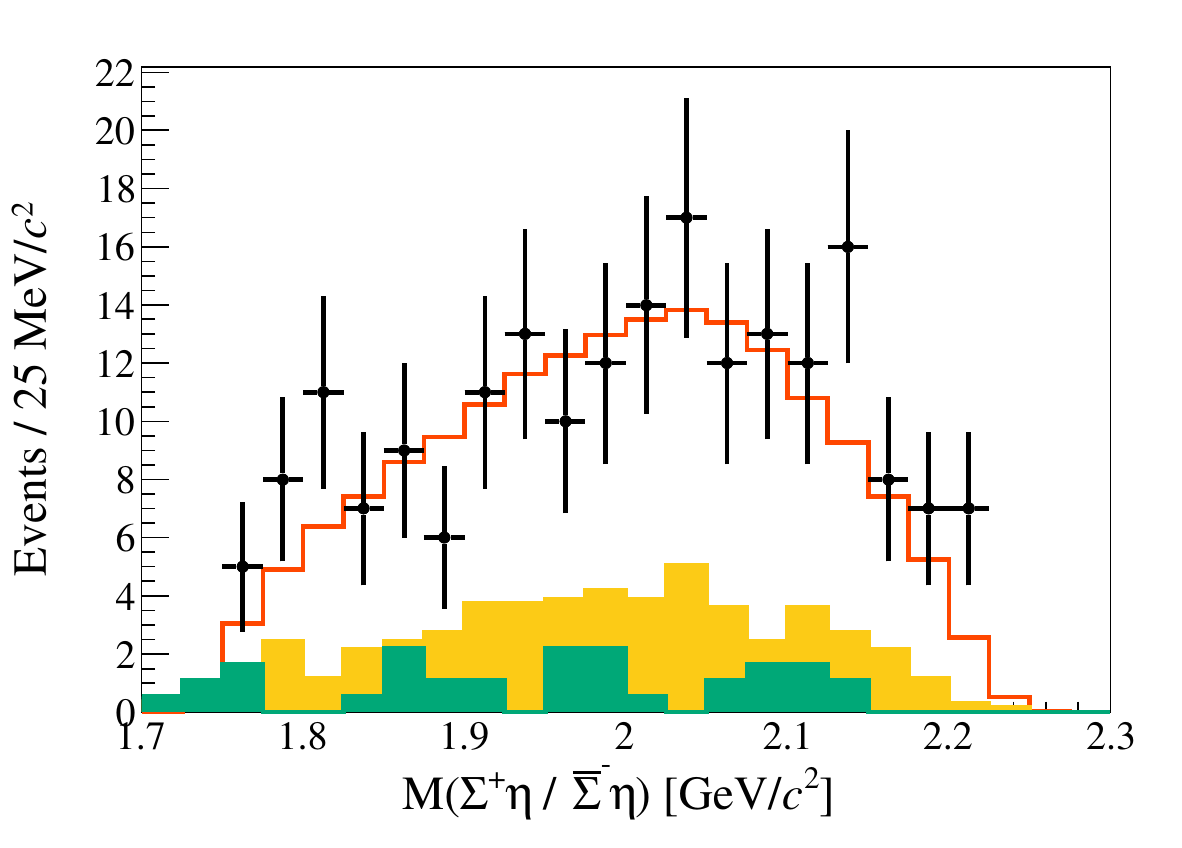}
\end{minipage}
\begin{minipage}[t]{0.49\linewidth}
\includegraphics[width=1\textwidth]{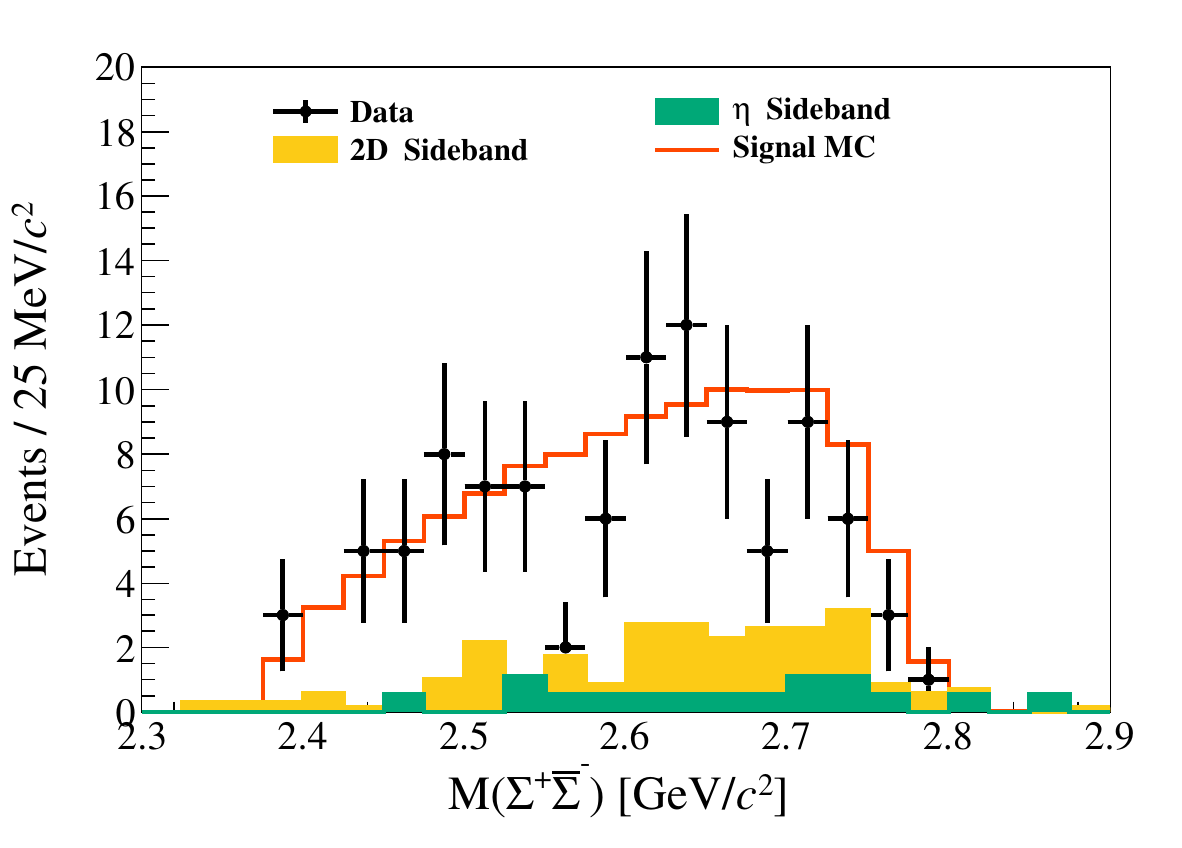}
\end{minipage}

\begin{minipage}[t]{0.49\linewidth}
\includegraphics[width=1\textwidth]{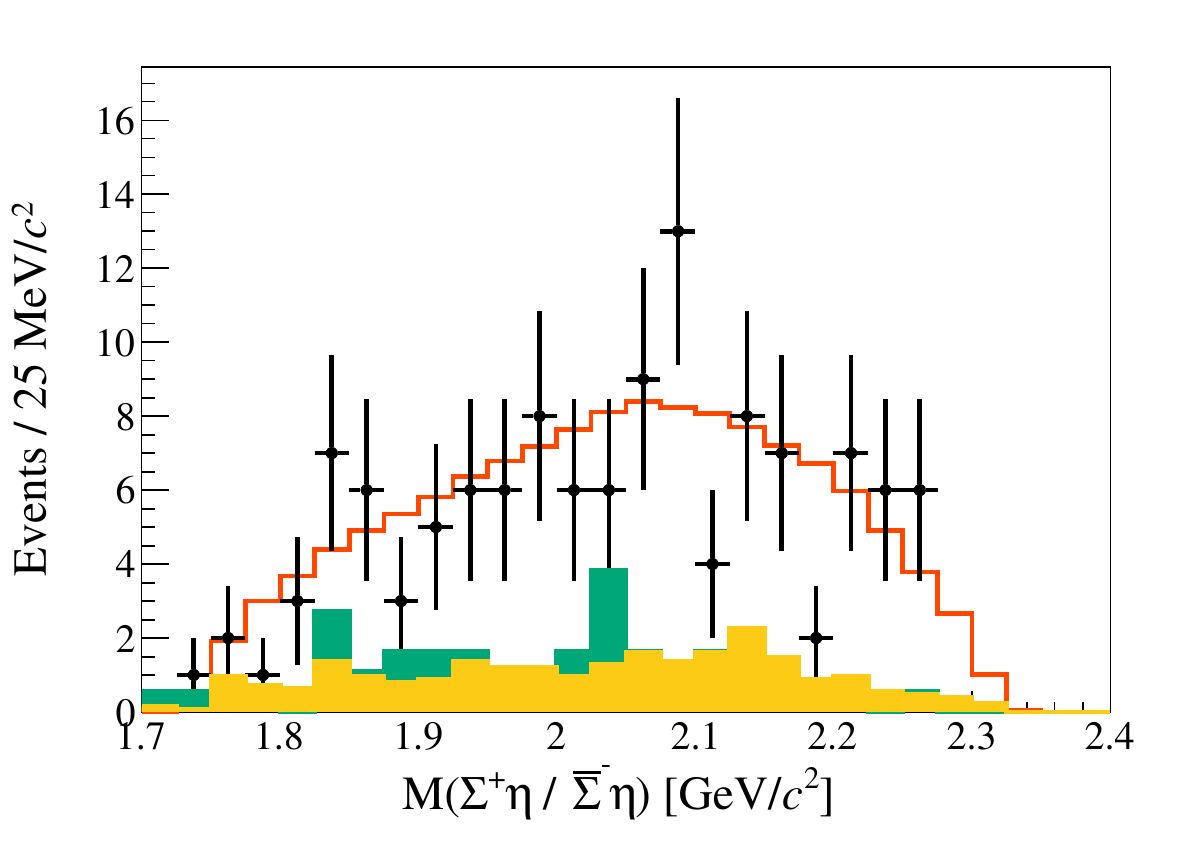}
\end{minipage}
\begin{minipage}[t]{0.49\linewidth}
\includegraphics[width=1\textwidth]{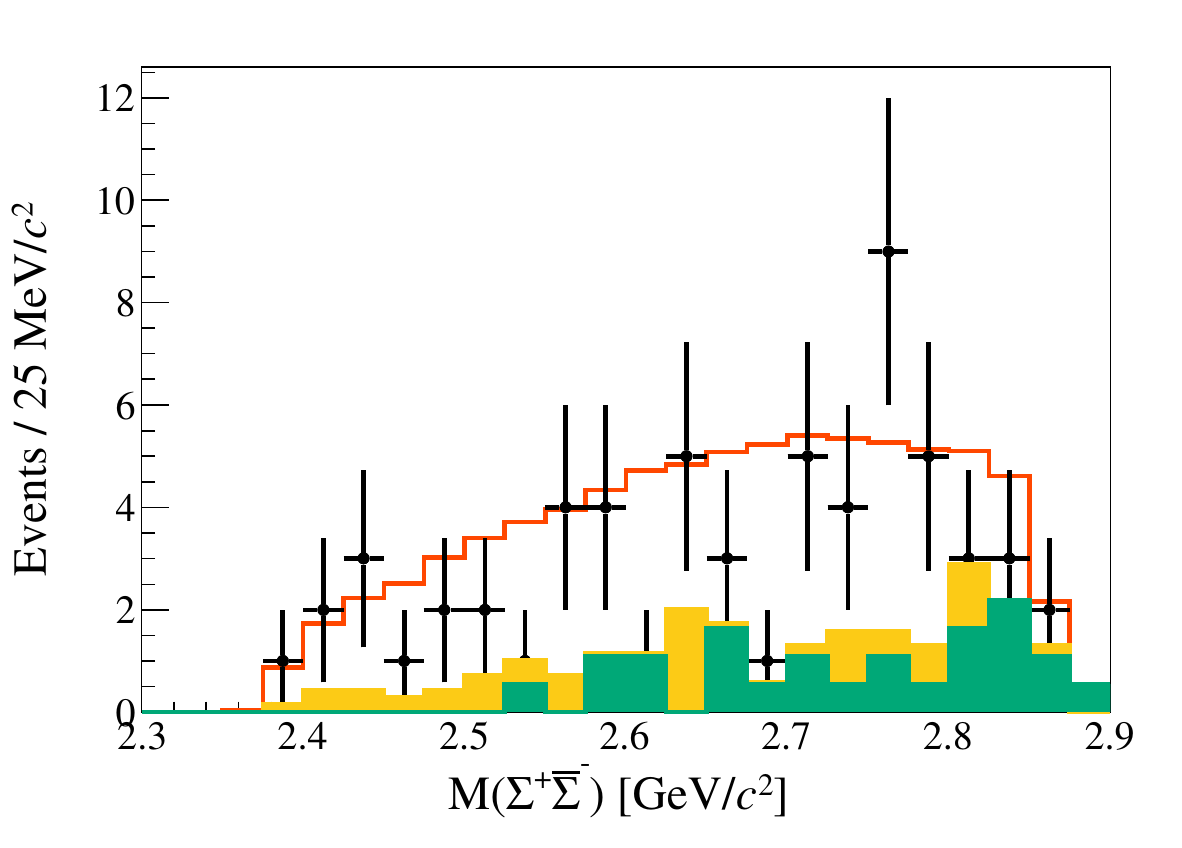}
\end{minipage}

\begin{minipage}[t]{0.49\linewidth}
\includegraphics[width=1\textwidth]{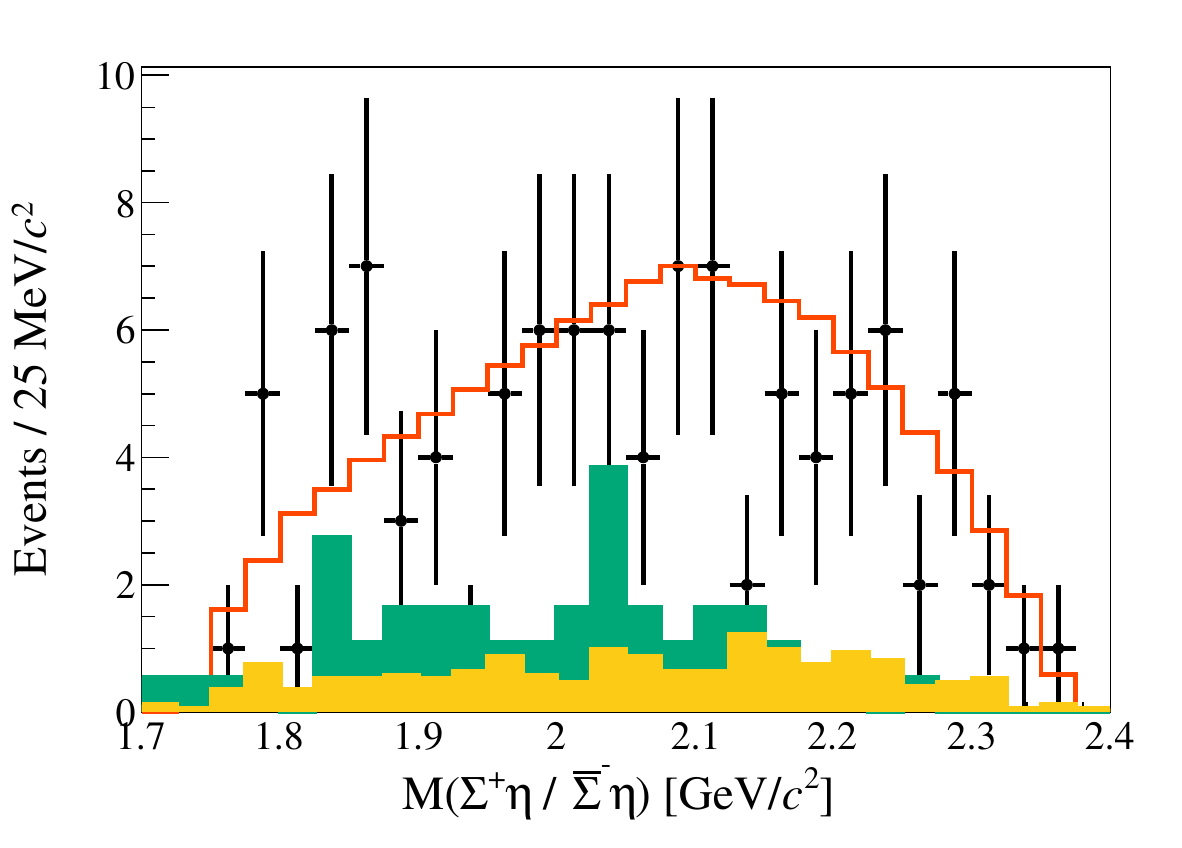}
\end{minipage}
\begin{minipage}[t]{0.49\linewidth}
\includegraphics[width=1\textwidth]{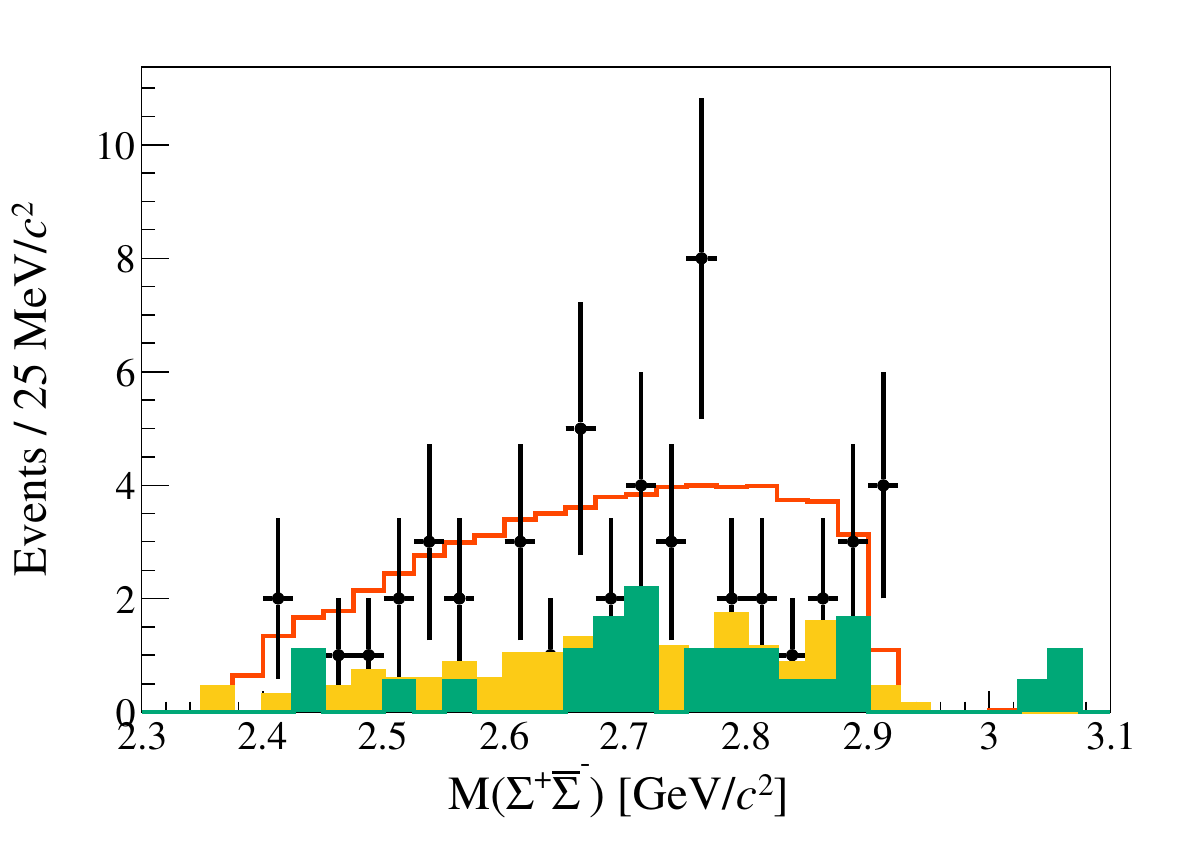}
\end{minipage}

\caption{Comparisons of $M(\bar{\Sigma}^{-}\eta /\Sigma^{+}\eta)$ and
$M(\Sigma^{+}\bar{\Sigma}^{-})$ of (top) $\chi_{c0}$, (middle)
$\chi_{c1}$, (bottom) $\chi_{c2}$, between (left) the data and (right)
individual BODY3 signal MC samples with all event selection
criteria. } \label{BODY3_hists} \end{center} \end{figure*}

The background subtracted $M(\Sigma^{+}\bar{\Sigma}^{-})$, $M(\Sigma^{+}\eta)$, and
$M(\bar{\Sigma}^{-}\eta)$ distributions of data 
are examined for possible intermediate structures, and no
obvious structure is observed with the current statistics.  The
data-MC comparison is shown in Fig.~\ref{BODY3_hists}.

\section{Systematic Uncertainties}\label{sec:sysU}

The systematic uncertainties considered are from the tracking and PID
efficiencies, photon reconstruction, kinematic fit, mass windows, mass
vetoes, fit method, scale factor $f_{\eta}$,
non-$\Sigma^{+}\bar{\Sigma}^{-}$ sideband background level, possible
intermediate states, external branching fractions, number of
$\psi(3686)$ events, and MC statistics. Each of these uncertainties is
discussed in detail below.

\begin{itemize}

\item{{\bf Tracking}}: The systematic uncertainties due to the
  tracking are estimated to be $0.7\%$ for $p$ and $1.0\%$ for
  $\bar{p}$~\cite{Ablikim:2011kv}. Adding them linearly
  gives the total systematic uncertainty due to $p$ and $\bar p$
  tracking to
  be 1.7\%.

\item{\boldmath \bf $p\bar p$ PID:} The systematic uncertainty due
  to the PID efficiency is estimated to be $0.5\%$ and $0.6\%$ for the
  proton and anti-proton~\cite{syspi}, respectively. Adding them
  linearly gives a total systematic uncertainty due to $p$ and $\bar p$ PID
  of 1.1\%.

\item{{\bf Photon reconstruction}}: The systematic uncertainties due
  to the photon reconstruction, which is $0.5\%$ for each photon
  \cite{Ablikim:2010zn}, is estimated by using the control sample of
  $J/\psi \to \pi^{+}\pi^{-}\pi^{0}$. There are seven photons in the
  final states, and the total systematic uncertainty of photon
  reconstruction is assigned as $3.5\%$.

\item{\bf kinematic fit}: The systematic uncertainty associated with
  the 6C kinematic fit is assigned as the difference between the
  efficiencies before and after the helix correction \cite{HelixG},
  which are $0.8\%, 0.4\%$ and $0.4\%$ for $\chi_{c0, 1,
    2}\to\Sigma^{+}\bar{\Sigma}^{-}\eta$, respectively.

\item{\bf Mass windows}:
To estimate the systematic uncertainties due to the mass windows of
$\Sigma^{+}$, $\bar{\Sigma}^{-}$ and $\eta$, we use the control sample
of $\psi(3686)\to\Sigma^{+}\bar{\Sigma}^{-}$ and
$\psi(3686)\to\eta\phi$. By comparing the difference between data and
MC simulation, the systematic uncertainty due to the $\Sigma^{+}$ mass
window is found to be negligible for $\chi_{c0,1,2}$; the systematic
uncertainty due to the $\bar\Sigma^{-}$ mass window is assigned to be
$0.3\%$ for $\chi_{c0,1,2}$; and the systematic uncertainty due to the
$\eta$ mass window is assigned to be $1.1\%$ for $\chi_{c0,1,2}$.

\item{\bf Mass vetoes}:
To estimate the systematic uncertainties of the mass vetoes, we
examine the branching fractions after enlarging or shrinking the veto
region. For different background vetoes, we vary the corresponding
mass windows for seven times with a step of 2 or 10 MeV$/c^2$. For
each case, the deviation between the alternative and nominal fits is
defined as $\zeta =$ ${|\mathcal B_{\rm nominal}-\mathcal B_{\rm
    test}|}\over{\sqrt{|\sigma^{2}_{\mathcal B,\rm
      nominal}-\sigma^{2}_{\mathcal B,\rm test}|}}$, where ${\mathcal
  B}$ is the branching fractions of
$\chi_{cJ}\to\Sigma^{+}\bar{\Sigma}^{-}\eta$ and $\sigma$ is its
statistical uncertainty. If $\zeta$ is less than 2.0, the associated
systematic uncertainty is negligible according to the Barlow
test~\cite{BarlowRef}. The largest relative difference is assigned
as the systematic uncertainty. 

 \item{\bf Fit range}: The systematic uncertainties due to the fit
   range are examined by enlarging and shrinking the fit range seven
   times with a step of 4 MeV$/c^{2}$, and the Barlow test is
   performed with the same method mentioned above, and the systematic
   uncertainties are negligible for $\chi_{0,1,2}$.

 \item{\bf Signal shape}: The systematic uncertainty arising from the
   signal shape is evaluated by comparing the fitted results obtained
   from the simulated signal shape before and after it is convolved
   with a Gaussian function. The differences in the measured branching
   fractions are taken as the systematic uncertainties, which is
   negligible for $\chi_{c0}\to\Sigma^{+}\bar{\Sigma}^{-}\eta$, and
   are 1.4\% and 0.4\% for
   $\chi_{c1,2}\to\Sigma^{+}\bar{\Sigma}^{-}\eta$, respectively.

\item{\bf Background shape}: The systematic uncertainty due to the
  background shape is estimated by replacing the second order
  Chebyshev polynomial function with a first or third order Chebyshev
  polynomial function. The largest differences in the measured
  branching fractions are taken as the systematic uncertainties, which
  are 6.3\%, 8.6\% and 2.9\% for $\chi_{c0, 1,
    2}\to\Sigma^{+}\bar{\Sigma}^{-}\eta$, respectively.

\item{\boldmath \bf Scale factor $f_\eta$}: The scale factor
  $f_{\eta}$ directly affects the fitted signal yields.  The
  associated systematic uncertainty is estimated by changing the
  $\eta$ sideband region by $\pm 1\sigma$, where $\sigma$ is the
  mass resolution of $\eta$. The largest differences of the branching
  fractions, which is negligible for
  $\chi_{c0}\to\Sigma^{+}\bar{\Sigma}^{-}\eta$, and are $3.5\%$ and
  $3.4\%$ for $\chi_{c1,2}\to\Sigma^{+}\bar{\Sigma}^{-}\eta$,
  respectively, are taken as the systematic uncertainties.

\item{\boldmath \bf Non-$\Sigma^{+}\bar{\Sigma}^{-}$ background
  level}: The systematic uncertainties induced by the
  non-$\Sigma^{+}\bar{\Sigma}^{-}$ background subtraction are
  estimated by changing the number of background events in the
  $\Sigma^{+}\bar{\Sigma}^{-}$ sideband region by $\pm 1\sigma$, where
  $\sigma$ is the statistical uncertainty of background
  events. The differences between the nominal and adjusted values
  are assigned as the systematic uncertainties, which are $4.8\%,
  6.3\%$ and $3.3\%$ for
  $\chi_{c0,1,2}\to\Sigma^{+}\bar{\Sigma}^{-}\eta$, respectively.

\item{\bf Possible intermediate states}: Considering the difference between data and PHSP signal MC sample, we develop a data-driven BODY3 model~\cite{BODY3}. The Dalitz plot of $M^{2}_{\Sigma^{+}\eta}$ versus $M^{2}_{\bar \Sigma^-\eta}$ obtained from data is taken as input for the BODY3 model, which is corrected for backgrounds and efficiencies. To estimate the systematic uncertainty associated with the possible intermediate structures, we compare the efficiencies based on the PHSP and BODY3 signal MC samples. The differences are assigned as the systematic uncertainties, which are $2.1\%, 3.6\%$ and $3.9\%$ for $\chi_{c0,1,2}\to\Sigma^{+}\bar{\Sigma}^{-}\eta$, respectively.
    
\item{\bf External branching fractions}: The branching fractions of
  $\psi(3686)\to\gamma\chi_{cJ}, \Sigma^{+}\to p \pi^{0},
  \bar{\Sigma}^{-}\to\bar{p}\pi^{0}, \eta\to\gamma\gamma$, and
  $\pi^{0}\to\gamma\gamma$ are taken from the PDG~\cite{pdg}. Their
  uncertainties are $2.4\%, 2.8\%$ and $2.4\%$ for $\chi_{c0, 1,
    2}\to\Sigma^{+}\bar{\Sigma}^{-}\eta$, respectively.

\item{\bf Number of $\psi(3686)$ events}: The number of $\psi(3686)$
  events is determined with the inclusive hadronic $\psi(3686)$ decays,
  and its uncertainty is assigned as 0.5\%~\cite{Ablikim:2017wyh}.

\item{\bf MC statistics}: Using simulated signal events of all the
  decay modes, the statistical uncertainty in the efficiency is
  $\Delta_{\epsilon}=\sqrt{\epsilon(1-\epsilon)/N}$, where $\epsilon$
  is the reconstruction efficiency after all the event selection, and
  N is the number of generated events.  The uncertainties for the MC
  statistics, estimated as $\Delta_{\epsilon}/\epsilon$ are 0.4\%,
  0.4\% and 0.4\% for $\chi_{c0,1,2}$, respectively.

\end{itemize}
\newpage
\begin{table}[htp]
\begin{center}
\caption{The relative systematic uncertainties (in unit of \%) in the measurments of the branching fractions. }
\label{list_sys}
\begin{tabular}{l  c  c  c}
\hline \hline  Source & $\chi_{c 0}$ & $\chi_{c 1}$ & $\chi_{c 2}$ \\
\hline   
Tracking   & $1.7$  & $1.7$  & $1.7$  \\
$p(\bar{p})$ PID   & $1.1$ & $1.1$  & $1.1$  \\
Photon reconstruction&  $3.5$    &   $3.5$   &  $3.5$   \\
Kinematic fit &  0.8 &  0.4  &  0.4  \\
 $\Sigma^{+}$ mass window & -- & -- & -- \\ 
 $\bar{\Sigma}^{-}$ mass window & 0.2 & 0.2 & 0.2\\
 $\eta$ mass window & 1.1 & 1.1 & 1.1 \\
Veto $\pi^{0}(\gamma_{1}\gamma_{E})$ & --  & -- & --  \\
Veto $\pi^{0}(\gamma_{2}\gamma_{E})$  & --  & -- & --  \\
Veto $J/\psi$  & $3.3$  &  --  &  --  \\

Fit range  & {--} & {--}  & {--}  \\
Signal shape & {--}  & {1.4}  & 0.4  \\
Background shape & {6.3}  & {8.6}  & 2.9  \\

Scale factor $f_{\eta}$ & -- & 3.5 & 4.4 \\
Non-$\Sigma^{+}\bar{\Sigma}^{-}$ background level & 4.8 & 6.3 & 3.3 \\
Possible intermediate states & { 2.1 }  & { 3.6 }  & { 3.9 }\\

External branching fractions &  2.4 &  2.8  &  2.4 \\
Total number of $\psi(3686)$ events & {{1.0}}  & {{1.0}}  & {{1.0}}  \\
MC statistics & 0.4 & 0.4 & 0.4\\
\hline Total  & {10.4}  & {{13.2}}  & {{9.2}}  \\
\hline\hline 
\end{tabular}
\end{center}
\end{table}
 
The systematic sources and their contributions are summarized in Table~\ref{list_sys}. The total systematic uncertainty for each signal decay is obtained by adding all of them in quadrature.

	
\section{Summary}\label{sec:summary}
	
By using $(27.12\pm0.14)\times 10^8$ $\psi(3686)$ events taken by the
BESIII detector, the decay
$\chi_{c0}\to\Sigma^{+}\bar{\Sigma}^{-}\eta$ is observed for the first
time with a statistical significance of $7.0\sigma$. Evidence for
$\chi_{c1}\to\Sigma^{+}\bar{\Sigma}^{-}\eta$ ($\chi_{c2}\to\Sigma^{+}\bar{\Sigma}^{-}\eta$)
is found with statistical significance of $4.3\sigma$ ($4.6\sigma$).
The branching fractions of these decays are determined to be
$\Br\left(\chi_{c 0} \rightarrow
\Sigma^{+}\bar{\Sigma}^{-}\eta\right)=(1.26 \pm 0.20 \pm 0.13)\times
10^{-4}, \Br\left(\chi_{c 1} \rightarrow
\Sigma^{+}\bar{\Sigma}^{-}\eta\right)=(5.10 \pm 1.21 \pm 0.67)\times
10^{-5}$ and $\Br\left(\chi_{c 2} \rightarrow
\Sigma^{+}\bar{\Sigma}^{-}\eta\right)=(5.46 \pm 1.18 \pm 0.50)\times
10^{-5}$, where the first and second uncertainties are statistical and
systematic, respectively. With current statistical precision, no
obvious intermediate structure is observed in these decays. In order to further understand the characteristics of $\chi_{cJ}$ mesons, the theoritical study of these decay channles is on the top toe for expectation.

\section{Acknowledgement}

The BESIII Collaboration thanks the staff of BEPCII and the IHEP computing center for their strong support. This work is supported in part by National Key R\&D Program of China under Contracts Nos. 2020YFA0406300, 2020YFA0406400; National Natural Science Foundation of China (NSFC) under Contracts Nos. 11635010, 11735014, 11835012, 11935015, 11935016, 11935018, 11961141012, 12025502, 12035009, 12035013, 12061131003, 12192260, 12192261, 12192262, 12192263, 12192264, 12192265, 12221005, 12225509, 12235017, 12150004; Program of Science and Technology Development Plan of Jilin Province of China under Contract No.20210508047RQ and 20230101021JC; the Chinese Academy of Sciences (CAS) Large-Scale Scientific Facility Program; the CAS Center for Excellence in Particle Physics (CCEPP); Joint Large-Scale Scientific Facility Funds of the NSFC and CAS under Contract No. U1832207; CAS Key Research Program of Frontier Sciences under Contracts Nos. QYZDJ-SSW-SLH003, QYZDJ-SSW-SLH040; 100 Talents Program of CAS; The Institute of Nuclear and Particle Physics (INPAC) and Shanghai Key Laboratory for Particle Physics and Cosmology; European Union's Horizon 2020 research and innovation programme under Marie Sklodowska-Curie grant agreement under Contract No. 894790; German Research Foundation DFG under Contracts Nos. 455635585, Collaborative Research Center CRC 1044, FOR5327, GRK 2149; Istituto Nazionale di Fisica Nucleare, Italy; Ministry of Development of Turkey under Contract No. DPT2006K-120470; National Research Foundation of Korea under Contract No. NRF-2022R1A2C1092335; National Science and Technology fund of Mongolia; National Science Research and Innovation Fund (NSRF) via the Program Management Unit for Human Resources \& Institutional Development, Research and Innovation of Thailand under Contract No. B16F640076; Polish National Science Centre under Contract No. 2019/35/O/ST2/02907; The Swedish Research Council; U. S. Department of Energy under Contract No. DE-FG02-05ER41374.


	

\end{document}